\documentclass[twocolumn,amsmath,amssymb,prb,longbibliography]
{revtex4-2}
\usepackage{xcolor}
\usepackage{multirow}
\usepackage{array}
\usepackage{amsmath}
\usepackage{graphicx}
\usepackage{dcolumn}
\usepackage{bm}
\usepackage{verbatim}
\DeclareMathAlphabet \mathbfcal{OMS}{cmsy}{b}{n}

\begin{document}

\title{Discernible signatures of fractionally charged anyons in a Pfaffian-Laughlin state}
\author{ Vadym Apalkov$^a$}
\author{ Tapash Chakraborty$^b$}

\affiliation{$^a$Department of Physics and Astronomy, Georgia State University, Atlanta, Georgia,30303, USA, $^b$ Department of Physics and Astronomy, University of Manitoba, Winnipeg, MB, Canada}

\date{\today}

\begin{abstract}
Understanding the nature of quasihole excitations, i.e., anyons that have
fractional charge and fractional statistics, has been a challenging problem in condensed matter physics. Our
theoretical approach to this problem has been to consider a quantum dot containing a few charged
particles, Coulomb coupled to the incompressible fluid. The size of the dot and the separation distance of the dot from the incompressible state strongly influences the energetics of the Laughlin quasiholes. Photoluminescence (PL) spectroscopy studies of this system have previously been
able to probe these quasiholes that have confirmed our estimates of the Laughlin quasihole energies. Turning to the Pfaffian state,
we now observe that such a theoretical approach is also able to provide valuable information about the Pfaffian
quasiholes, viz., the energy dispersion, the charge density distribution and the quasihole energy.
The energy dispersion of e/4 quasiholes derived here, clearly reflect the interaction between the
quantum dot and the incompressible Pfaffian state.  PL spectroscopy experiments on the 5/2
Pfaffian-Laughlin state could perhaps shed light on the energetics we found here of  these elusive quasiparticles.

\end{abstract}

\maketitle

In recent intense studies of nanoscale quantum materials, symmetry and 
topology have been front and center in condensed matter physics\cite{Tkachov_2022,40_years_QHE,
Encyclopedia,Gruber_review_topological_2025,
Review_tapash_Physica_E_2026}. 
Interactions among constituent particles in monolayer and more importantly in multilayer  
Dirac materials hosting Dirac fermions (one of the most comprehensively studied quantum
materials) have been responsible for a variety of unique properties that are well 
documented in the literature\cite{Abergel_long_range_PRL_2009,
Abergel_PRB_2009,Berashevich_PRB_2011,Berashevich_J_phys_chem_2011,
Traits_characteristics_interacting_Dirac_fermions,
My_gap_structure_Hofstadter_2014,
Apalkov_fractal_Butterflies_2015,Apalkov_trilayer_PRB_2012}.   
The most prominent example is, of course, the presence of the incompressible Laughlin 
state\cite{Two_dimensional_electron_correlation,Eisenstein_science_1990,
Anomalous_Quantum_Hall_Effect,
Book_FQHE,Chakraborty_PRB_1985,
Half_polarized_PRL_2001,
Evidence_phase_transition_FQHE} in these systems\cite{FQHE_graphene,phase_transitions_bilayer,
Tunability_FQHE_dirac_materials_PRB_2014}.

One of the other most exciting prospects in this regard is the possibility of the existence
of Pfaffian states in bilayer graphene\cite{stable_pfaffian_bilayer,
Interacting_fermions_book,Wenchen_PRB_2017}.
It has long been argued that the Landau level filling factor 
$\nu=5/2$\cite{Observation_willett_1987} harbors the non-Abelian fractional quantum Hall (FQH) liquid -- one of the most exotic 
quantum materials that, due to robustness of exchange statistics against local noise, could make anyons in this system a reliable platform for topological quantum computation \cite{Ma_encyclopedia,Topological_quantum_computing_rowell}.
The fractional charge\cite{resonant_tunneling_QHE_1995,
Obervation_Laughlin_quasiparticles_PRL_1997}  and the fractional statistics\cite{Anyon_collision,2d_anyon_gas}, the defining characteristics of the Laughlin quasiparticle excitations have been observed earlier. However, the nature of the Pfaffian quasiparticle excitations still remains rather elusive.

The FQH effect in even-denominator filling factors has already been detected 
experimentally in multilayer graphene\cite{Trialayer_graphene_tunable_even_2024,trilayer_observation_2025,
sha2025cascadeevendenominatorfractionalquantum}.
The challenge has been how to associate these even-denominator FQH observations
with the Pfaffian states\cite{Paired_Hall_states_PRL_1991,
Paired_Hall_states_Wen_1992,Noneabelions_Read_1991}. 
In this pursuit, we explore here the energetics of the Pfaffian quasiholes. The Pfaffian wave function is the exact ground state of a singular three-body model interaction
\cite{Exact_solutions_and_the_adiabatic_1992,
Paired_Hall_states_PRL_1991} $V_{\text{pf}}=V_0 \sum_{i<j<k} P_{ijk}(L_{\text{max}})$, where $P_{ijk}(L_{\text{max}})$ is the projection operator onto a three-electron state with orbital angular momentum $L_{max}=3$. Here, $V^{}_0$ is a constant. Exact solutions for quasiholes (as shown below) are also available for this model interaction.
We have previously established\cite{stable_pfaffian_bilayer,
Interacting_fermions_book} that for bilayer 
graphene  with AB stacking, there are two special Landau levels: one in the K valley
and another one in the K' valley, for which there is a range of magnetic fields where the 
stable half-filled ground state is determined by the Pfaffian 
function\cite{Treatise_determinants} acting on the 
half-filled Laughlin state\cite{Anomalous_Quantum_Hall_Effect,Book_FQHE}. Stability of such a ground state, i.e., its 
collective excitation gap, depends on the magnetic field and for a finite magnetic field, 
the $\nu= 1/2$-FQHE state becomes even more stable than the corresponding state in a
conventional electron system. Another important property of bilayer graphene is that 
there are external parameters, such as the bias voltage and the in-plane magnetic field 
that can alter the inter-electron interaction strength within a given Landau level and
correspondingly change the properties of the $\nu= 1/2$ Pfaffian state.

Before we report on our work on the Pfaffian quasiholes, let us briefly describe our previous theoretical studies of the Laughlin quasiholes. We previously developed a theoretical approach to analyze the energetics of fractionally-charged
Laughlin quasiholes at the Laughlin state ($\nu=\frac{1}{3}$)\cite{QD_FQHE_my_2002,
spin_transitions_my_PRB_2006}, where a parabolic quantum dot (QD)\cite{Chakraborty_QDs,Maksym_QDs_magnetic_field_1990}
was coupled, via {\it only} the Coulomb force, to a two-dimensional electron system which is in a Laughlin state.
There we found that, in the case of a single electron in the dot the physics is somewhat similar to that
of a point impurity in a Laughlin state investigated earlier\cite{Rezayi_Haldane_1985}. In this case, the QD emits a fractionally
charged quasihole (e=1/3) that orbits around the QD, as evidenced from the charge-density calculations\cite{QD_FQHE_my_2002,
spin_transitions_my_PRB_2006,Rezayi_Haldane_1985}.
However, we found that for two interacting electrons in the quantum dot, the collective excitation exhibits
an oscillatory behavior which is due to confinement of the fractionally-charged quasihole excitations by
the quantum dot. This is purely a consequence of interelectron interactions in the dot. 

Interestingly,
for a charge-neutral dot there is no dispersion of the energy as a function of the angular momentum,
and most importantly, the incompressible liquid is not influenced by the dot at all\cite{QD_FQHE_my_2002,
spin_transitions_my_PRB_2006,QD_FQHE_PRL_2003}. On the other hand,
the energy of the qd-liquid is significantly lowered for a charged QD, as compared to the isolated QD or
the incompressible liquid without the dot. Calculation of the charge density distribution have indicated
that the low-lying excited states of the qd-liquid can be described by the ionization process\cite{QD_FQHE_my_2002,
spin_transitions_my_PRB_2006,Rezayi_Haldane_1985} as emission
of a quasihole: If the net charge of a QD is negative, the ground state of the qd-liquid can be considered
as a QD plus three quasiholes. If we increase the angular momentum of the two-dimensional (2D) electrons, one of the quasiholes moves
away from the QD. This is inferred from the calculated charge distribution of electrons around the QD,
where a local minimum corresponds to the quasihole moving away from the QD as the angular momentum
is increased. The position of the local minimum at different angular momenta of the charge
density correspond to the orbit radius of the quasihole. 

Confinement of the quasihole in the incompressible
state by the quantum dot in close proximity is purely due to the interaction between the electrons in the quantum dot which
results in the specific charge distribution in the quantum dot and an additional interaction of a qusihole of
the incompressible liquid with the local excitation of the dot. In those studies\cite{QD_FQHE_my_2002,
spin_transitions_my_PRB_2006,
QD_FQHE_PRL_2003,QD_FQHE_exp_PhysicaE_2004,
charged_excitons_2004}, we also evaluated the quasihole energy 
(i.e., the energy of the quasihole wave function\cite{Anomalous_Quantum_Hall_Effect,
Chakraborty_PRB_1985,Morf_Halperin_PRB_1986}) for the QD-liquid complex that
is close to the theoretically obtained results.
Similar studies were also performed to investigate the ground state and
low-energy excitations of a quantum Hall liquid at a filling
factor $\nu=2/5$ which was Coulomb-coupled to a quantum dot (a QD-liquid complex) separated from the liquid by a distance
$d$. The dot consisted of two or three electrons and a hole. It
was found that the presence of the dot changed locally the
spin polarization of the electron liquid that depends crucially
on the separation distance\cite{spin_transitions_my_PRB_2006}.

Photoluminescence (PL) spectroscopy that has been an
useful tool to probe the anyonic properties\cite{anyons_bosonic_properties_2021} provides excellent support to our theoretical model\cite{QD_FQHE_PRL_2003,
QD_FQHE_exp_PhysicaE_2004,charged_excitons_2004}. 
The $\nu=\frac13$ fractionally charged quasiholes were probed with PL spectroscopy of a 2D electron gas confined in a quantum well and subjected to a perpendicular magnetic field. There the electron-hole pairs were excited with a single laser line, and then the electrons and holes relax and form excitons. The luminescence lines (as the electrons and holes recombine), when the lowest Landau level is fully occupied, show a nearly linear magnetic field dispersion.
This can be attributed to recombination of electrons from the occupied Landau levels with photocreated holes in the valence band. However, Sch\"uller et al.\cite{QD_FQHE_PRL_2003} observed that under certain circulstances, near the 1/3 FQHE state, the PL lines exhibit a significant reduction in energy at low temperatures compared to those at a much higher temperature where the lines are linear. Further, it was noted that a very small thermal energy was required to destroy that anomaly. Our calculated quasihole energy gap of the QD-liquid complex is precisely the energy that can be attributed to that energy anomaly.
Interestingly, those ideas have been revived recently in theoretical\cite{anyons_impurities_PRRes_2026,
Anyons_trions_PRX_2026} and experimental\cite{Signature_FQHE_anyons_trions_TMDC_2026} studies.

In contrast to the Laughlin state, where the fractionally-charged anyons that obey fractional
statistics\cite{Halperin_statistics_PRL_1984,Fractional_statistics_PRL_1984,
2d_anyon_gas} and have received a lot of attention, very little is known about the energetics
of the elementary excitations in the Pfaffian state\cite{Fradkin_Nuclear_1998}. Here, we demonstrate that our Coulomb-coupled QD and the Pfaffian-Laughlin system indeed provide invaluable information, not least about the energetics of the fractionally-charged Pfaffian anyons that could perhaps be observed experimentally, as it was the case for the Laughlin state.

Our Pfaffian-Laughlin state at Landau level filling fraction $\nu=\frac12$ has the form\cite{Paired_Hall_states_PRL_1991,
Paired_Hall_states_Wen_1992,Noneabelions_Read_1991}
\begin{equation}
\Psi^{}_{Pf}=\text{Pf}\left(\frac{1}{z^{}_i-z^{}_j}\right)
\Psi^{1/2}_L
\end{equation}
where $\text{Pf}\left(\frac{1}{z^{}_i-z^{}_j}\right)={\cal A}\prod_{i \in \text{even}}^N\frac{1}{z^{}_{i-1}-z^{}_j}$ is the
antisymmetrized product over pairs of electrons. A more formal definition of the Pfaffian is: The
determinant of a $N\times N$ antisymmetric matrix is the square of a polynomial, the Pfaffian\cite{Treatise_determinants}.
The Laughlin wave function for bosons in a half-filled Landau level $\Psi^{1/2}_L$ on the other hand, is the square of the
Vandermonde determinant -- a perfect square, and has been described in the literature as a correlated $d$-wave Bose fluid\cite{Bose_liquid_2d_2007}.

The Pfaffian-Laughlin state $\Psi^{}_{Pf}$ is a pairing state at one-half filling fraction that is incompressible and
obey non-Abelian statistics\cite{Paired_Hall_states_PRL_1991}. As charge is removed from the Pfaffian-Laughlin state (1), the fractionally
charged quasiholes are created. Following Laughlin \cite{Anomalous_Quantum_Hall_Effect,Book_FQHE}, the quasihole wave function is written
\begin{equation}
\Psi^{}_{qh}(z)=\prod^{}_k\left(z^{}_k-\eta\right)\, \text{Pf}\left(\frac1{z^{}_i-z^{}_j}\right)\prod_{i<j}\left(z^{}_i-z^{}_j\right)^2
\label{QH1}
\end{equation}
where $\eta$ is the quasihole position, and the resulting object will have charge $e/2$. Interestingly, for the Pfaffian quasihole wave function (\ref{QH1}), the quasihole can be split into two objects\cite{Fradkin_Nuclear_1998}
\begin{eqnarray}
\Psi^{}_{qh}(z)& = & \text{Pf}\left(\frac{\left(z^{}_i-\eta^{}_1\right)
\left(z^{}_j-\eta^{}_2\right)+\left(z^{}_j-\eta^{}_1\right)\left(z^{}_i-\eta^{}_2\right)}{z^{}_i-z^{}_j}\right) \nonumber \\
& & \times 
\prod_{i<j}\left(z^{}_i-z^{}_j\right)^2.
\end{eqnarray}
i.e., individual quasiholes will have charge $e/4$.  When the two quasiholes coincide $\left(\eta^{}_1=\eta^{}_2\right)$, we get back the original $e/2$ quasihole.
 Several experimental groups have reported the existence of $e/4$ quasiholes\cite{PNAS_quasiholes_2009,
 observation142026}.

Here we consider a liquid-QD complex that consists of two 2D layers separated by a distance $d$ along the $z$-direction,  where we allow strong Coulomb coupling
between them, while suppressing a direct interlayer
tunnelling\cite{Tapash_FQHE_1987,Interlayer_FQHE}.
The system is placed in an external magnetic field $B$ and in one of the two layers the Pfaffian state is formed, i.e., the lowest Landau level is completely occupied and the filling factor of the second Landau level is at $\nu =\frac{1}{2}$. In the following, we assume that the electrons are spinless, i.e., the Zeeman energy is large enough that the electron states are spin-polarized. In the second layer the charged particles are subjected to a parabolic confinement potential of the following form 
\begin{equation}
V_{\rm conf}(x,y)=\frac12m^*\omega_0^2\left(x^2+y^2\right),
\end{equation}
where the potential is characterized by the potential strength $\omega_0$ and the corresponding oscillator length is $l_{\rm dot}=\left(
\hbar/m^*\omega_0\right)^{\frac12}$. In all our calculations below we assume that 
$l_{\rm dot} = 15 $ nm. 
The second layer is therefore, effectively, a parabolic 2D quantum dot\cite{Chakraborty_QDs,Maksym_QDs_magnetic_field_1990,
Fock_Darwin_graphene_2007}.
Below we consider the two cases: (i) the quantum dot is occupied by one electron and (ii) the quantum dot is occupied by an exciton, i.e, a pair of electron and hole. 
The single particle wave functions of the quantum dot placed in perpendicular magnetic field can be written as
\begin{eqnarray*}
\psi_{n,l}(x,\phi) &= &
  \left(\frac{b}{2\pi l_0^2}\frac{n!}{(n+|l|)!}\right)^{1/2} \\
& & \times\sum_{j=0}^n C(n,l,j) e^{-il\phi} e^{-(x^2/2)}
 x^{2j+|l|}
\end{eqnarray*}
where $x=(b/2l_0)^{\frac12}r$, $b=(1+4\omega_0^2/\omega_c^2)^{\frac12}$, $l_0=\sqrt{\hbar}{|e|B}$ is the magnetic length, $B=15$ T is the magnetic field, $r$ and $\phi$ are polar coordinates, 
and
$$C(n,l,j)=(-1)^j\frac{(n+|l|)!}{(n-j)!(|l|+j)!j!},$$
where $n=0,1..$ is the radial quantum number and $l$ is the azimuthal 
quantum number. The corresponding single particle energy spectrum has the oscillator form 
$E_{l} = l \hbar [(\omega_c^2+\omega_0^2)^{1/2}-\omega_c]/2$. To make the system finite, we consider below 10 lowest single-particle quantum dot states.  

The interaction between all particles, both within each layer and between the layers, is the Coulomb interaction, which is given by 
\begin{eqnarray*}
H_{\rm int} &=&\frac{1}{\varepsilon}
\sum_{i<j}\frac{q_i q_j}{|\vec{r}_{D,j}-
\vec{r}_{D,i}|}+\frac{e^2}\varepsilon\sum_{i<j}\frac1{|\vec{r}_{L,j}-
\vec{r}_{L,i}|} \\
 & &  + \frac{e}{\varepsilon}\sum_{i,j}\frac{q_j}{\left[d^2+|\vec{r}_{D,j} -
 \vec{r}_{L,i}|^2\right]^{\frac12}}
 \end{eqnarray*}
where $e$ is the electron charge, $q_i$ and $q_j$ are the charges of the particles in quantum dot ($q_i=e$ for an electron and $q_i=-e$ for a hole), $\epsilon$ is the dielectric constant, 
$\vec{r}_{D}$ and $\vec{r}_L$ are
the 2D radius-vectors corresponding to particles in the quantum dot and the 2D electron layer, respectively. 
In what follows, for electrons in the layer corresponding to the incompressible state, we 
use a spherical geometry\cite{Fano_spherical_geometry,Haldane_PRL_1983}, which correctly takes into account the azimuthal symmetry of the system. In this case, $\vec{r}_L$ is the radius-vector of an electron on a sphere of radius $R\sqrt{S} l_0$, where $2S$ is the number of magnetic flux quanta though the sphere\cite{Fano_spherical_geometry}.

\begin{figure}[b]
\includegraphics[width=0.8\columnwidth]{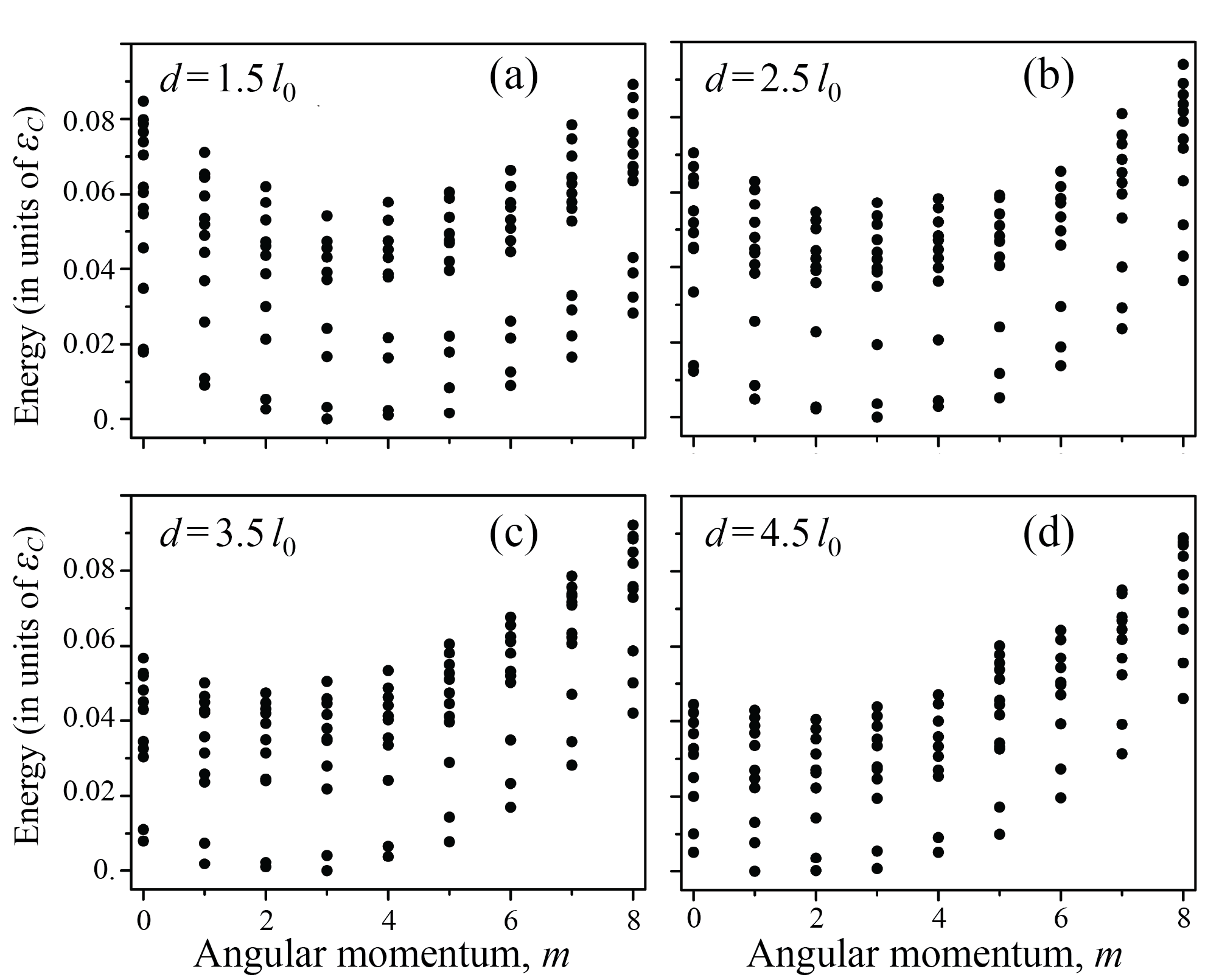}
\caption{\label{fig_1e_energy}  Energy spectra of a Pfaffian state coupled to a QD containing one electron. The distance between the 2D layer and the QD layer is $d= 1.5 l_0$ (a), 2.5 $l_0$ (b), 3.5 $l_0$ (c), 4.5 $l_0$ (d). Only a few low-energy states are shown for each angular momentum, $m$. } 
\end{figure}

For dealing with the 2D electron layer, we use the exact diagonalization method for a finite electron system with $N_L = 12$ electrons. The radius of the corresponding sphere is equal to $R=\sqrt{S} l_0$, where $2S = 21$. 
 Such a size of a sphere corresponds to the formation of the Pfaffian state, i.e., the filling factor of $\nu = \frac{1}{2}$. The number of single-particle states within the sphere is $2S+1 = 22$. 
The Coulomb interaction within a single layer is described in terms of Haldane pseudopotentials,   $V_m$\cite{Haldane_PRL_1983}, which are the energies of the interaction of two electrons with relative angular momentum $m$ and are given by the following expression 
\begin{equation}
V_m^{(n)} = 
 \int_0^{\infty } \frac{dq}{2\pi} qV(q) 
L_{1}^2(q^2/2) L_m (q^2) e^{-q^2},
\label{Vmm}
\end{equation}
where $V(q) =  \frac{e^2}{\epsilon l_0 q }$ is the Coulomb interaction in the momentum space,  $\epsilon$ is the dielectric constant, $L_n(x)$ are the Laguerre polynomials. The expression (\ref{Vmm}) takes into account that electrons occupy the second Landau level. The characteristic energy scale of the Coulomb inter-particle interaction is $\varepsilon_C = \frac{e^2}{\epsilon l_0 }$. All the energies reported here will be expressed in this energy unit.  

For a finite QD-liquid system, we perform the exact digonalization of the interaction Hamiltonian and obtain the 
many-electron energy spectra, $E_s$, with the corresponding wavefinctions, $\Phi_s (\vec{r}_{D,1},\ldots |\vec{r}_{L,1}
\ldots)$. Because of the azimuthal symmetry of the system, the states are characterized by the magnetic quantum number, $m$, which the $z$ component of the angular momentum. In addition to the energy spectra $E_i$, we also calculate the electron density distribution for electrons in the 2D electron layer. Such a density is determined from 
\begin{eqnarray}
\rho (r) & = &\int \cdots \int d\vec{r}_{D,1}\ldots
d\vec{r}_{L,1} \ldots \sum_{i=1}^{N_L} 
\delta (\vec{r}-\vec{r}_{L,i})  \nonumber \\
&  & \times\left|\Phi_s (\vec{r}_{D,1},\ldots |\vec{r}_{L,1}
\ldots)\right|^2
\end{eqnarray}

\begin{figure}[b]
\includegraphics[width=0.8\columnwidth]{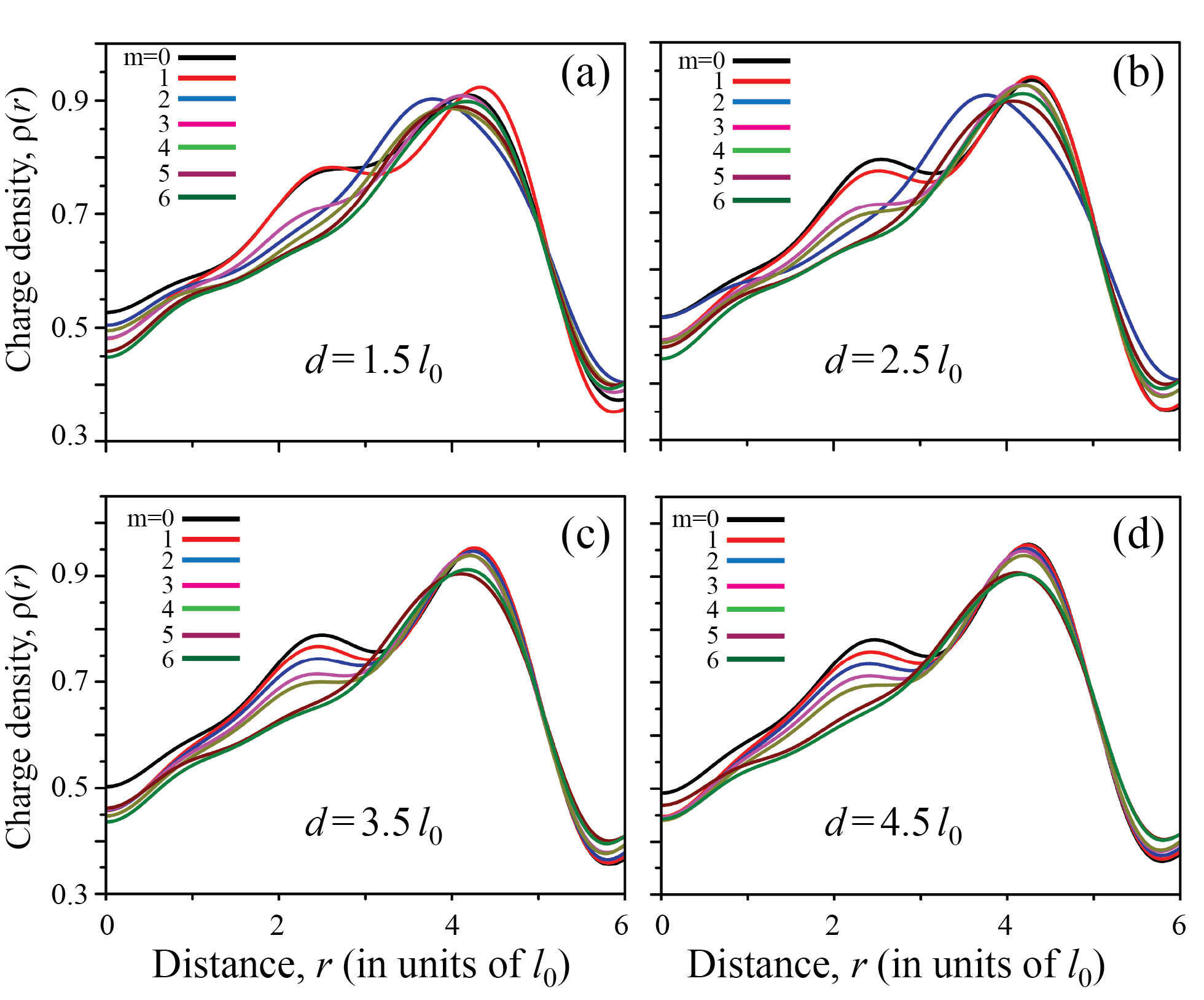}
\caption{\label{fig_1e_charge}  Electron density for a Pfaffian state coupled to a QD containing one electron. Different lines correspond to different values of the angular momentum of the system, as marked in each panel. For each angular momentum, $m$, the density is shown for the corresponding lowest state.  
The distance between the 2D layer and the QD layer is $d= 1.5 l_0$ (a), 2.5 $l_0$ (b), 3.5 $l_0$ (c), 4.5 $l_0$ (d).  } 
\end{figure}

Our results for the Pfaffian anyons, when the QD contains one electron, are presented in Figs. \ref{fig_1e_energy} and \ref{fig_1e_charge}. Clearly, in contrast to the Laughlin state, a confluence of factors is at work in the Pfaffian case. For the Laughlin state with a single electron in the QD\cite{QD_FQHE_my_2002,spin_transitions_my_PRB_2006}, we previously noticed that for a small separation $d \leq 1.5$, the perturbation of the incompressible liquid by the QD electron is strong, and the collective excitation is gapless. For a larger separation $d \approx 2.0 l_0$, there is a well-defined branch of excitations at \(m = 0\). 
However, in the present case of the Pfaffian-Laughlin state, for $d = 1.5 l_0$ and $d = 2.5 l_0$, we find clearly gapped excitations, and there are distinct branches of excitations for low momenta. For $d \geq 3.5$, only the doubly degenerate lower mode survives. The low-energy branch is characterized by the energy scale $\sim 0.02 \varepsilon_C$. For our $N_L = 12$ electron system, we calculated the  energy of one $e/2$ quasihole, which is introduced into the system by increasing the flux quanta by 1. Such an energy is $\Delta_{1/2} \approx 0.0271 \varepsilon_C$. Correspondingly, we can estimate the  energy of one $e/4$ quasihole as half of $\Delta _{1/2}$, giving $\Delta_{1/4} \approx \frac{1}{2}\Delta_{1/2} \approx 0.014 \varepsilon_C$. This energy scale falls within the characteristic energy range of the lowest branches clearly visible in Fig. \ref{fig_1e_energy}.  Incidentally, for the Laughlin quasihole, the energy is $0.0276 e^2/\epsilon\l^{}_0$\cite{Chakraborty_PRB_1985}.

The two low-energy branches can be attributed to the formation of two $e/4$-quasiholes, where the splitting of the branches is determined by quasihole-quasihole interactions. These two branches persist for all inter-layer separations, $d$. Furthermore, when the distance between the layers is small enough ($d\leq 2.5 l_0$), more well-defined branches become visible in the spectra. Specifically, for $d=1.5 l_0$ [Fig.\ \ref{fig_1e_energy}(a)] there are two additional well-formed low-energy branches at $m>0$, while for $d=2.5 l_0$, there is one extra well-defined low-energy branch. For such small distances, the perturbation of the system is relatively strong, which results in the excitation of many fractionally charged quasiholes. The energies of these extra branches are shifted from the lowest branch by $\approx 0.02 \varepsilon_C$. 

\begin{figure}[b]
\includegraphics[width=0.8\columnwidth]{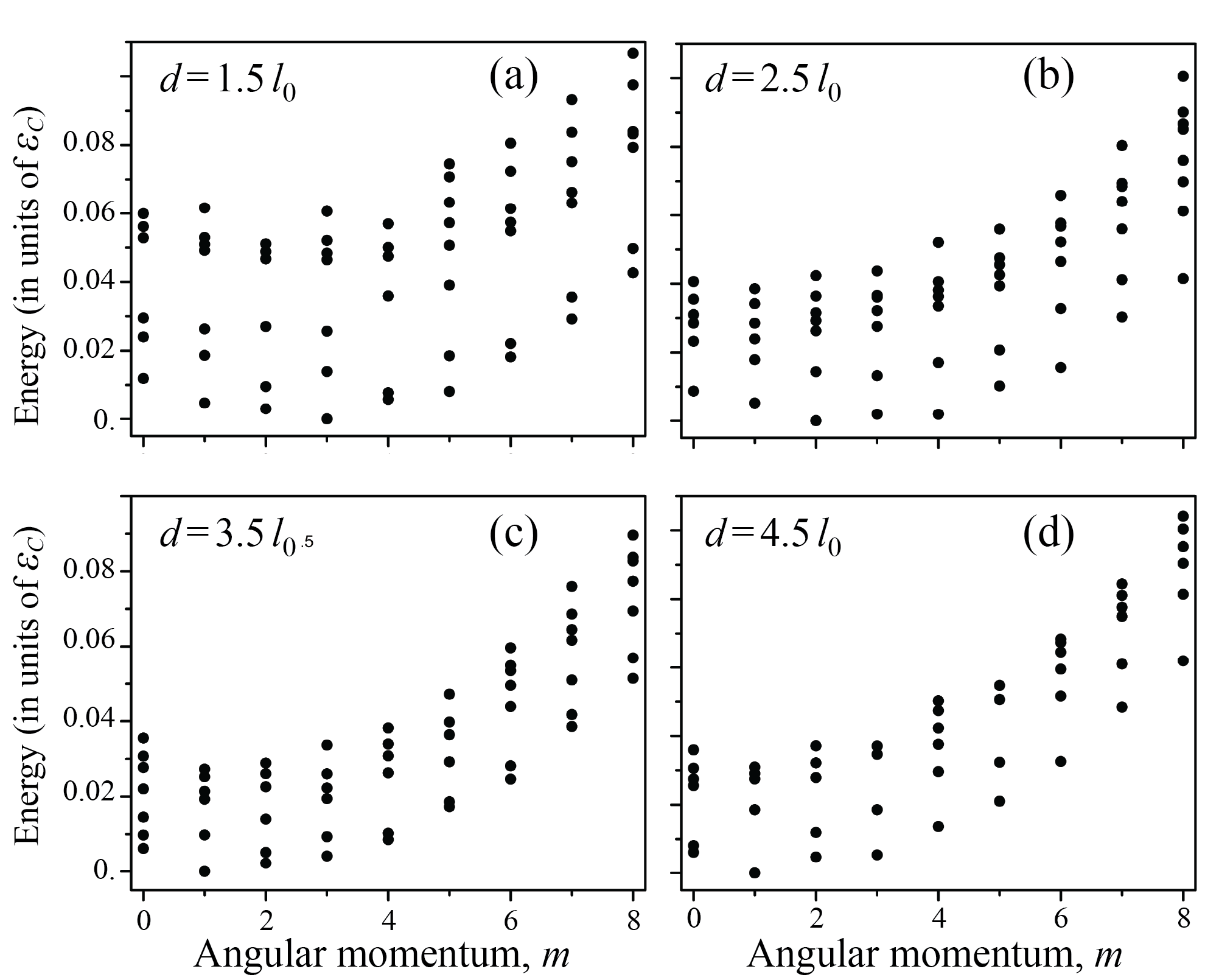}
\caption{\label{fig_1e1h_energy}  Energy spectra of a Pfaffian state coupled to a QD containing one electron and one hole. The distance between the 2D layer and the QD layer is $d= 1.5 l_0$ (a), 2.5 $l_0$ (b), 3.5 $l_0$ (c), 4.5 $l_0$ (d). Only a few low-energy states are shown for each angular momentum, $m$.  } 
\end{figure}

The charge distribution for the one-electron case is shown in Fig. \ref{fig_1e_charge} for a few lowest energy levels. Here again, we see a sharp contrast to the Laughlin state. In the latter case, the minima in the charge density were identified\cite{QD_FQHE_my_2002,
spin_transitions_my_PRB_2006,QD_FQHE_PRL_2003, QD_FQHE_exp_PhysicaE_2004,charged_excitons_2004} with the center of the quasiparticle defect (fractionally charged) emitted by the QD electron. For the distribution shown in Fig. \ref{fig_1e_charge} we see a different behavior. With increasing $m$ there is no clear shift of the position of the distribution maximum, which suggests that the structure of the low-energy states of the Pfaffian-QD complex is not a single excited Pfaffian mode but a bound state of many quasiholes. This is consistent with the energy spectra of the system shown in Fig.\ \ref{fig_1e_energy} where there are many almost degenerate low-energy branches.

\begin{figure}[b]
\includegraphics[width=0.8\columnwidth]{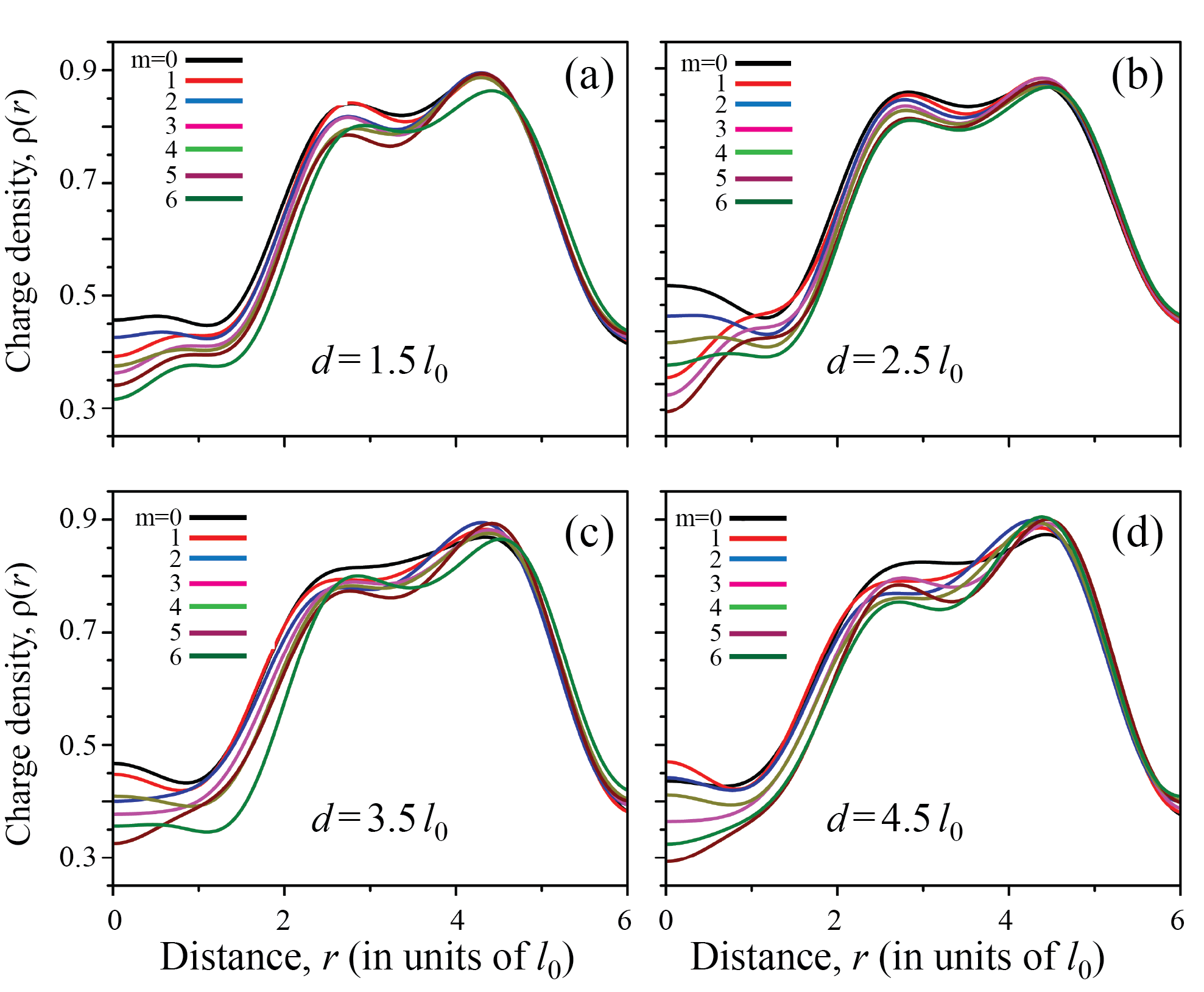}
\caption{\label{fig_1e1h_charge} Electron density for a Pfaffian state coupled to a QD containing one electron and one hole. Different lines correspond to different values of the angular momentum of the system, as marked in each panel. For each angular momentum, $m$, the density is shown for the corresponding lowest state.  
The distance between the 2D layer and the QD layer is $d= 1.5 l_0$ (a), 2.5 $l_0$ (b), 3.5 $l_0$ (c), 4.5 $l_0$ (d).   } 
\end{figure}

The results for the system with the QD occupied by an exciton are shown in Figs. \ref{fig_1e1h_energy} and \ref{fig_1e1h_charge}. The interaction of the QD with the Pfaffian state is now weaker and is of the dipole type. In this case, for all interlayer distances $d$, we observe the formation of only two low-energy branches. For large interlayer distances ($d = 4.5 l_0$, see Fig. \ref{fig_1e1h_energy}), the separation between the low-energy branches increases; consequently, the repulsion between quasiholes becomes more pronounced as the interaction between the QD and the 2D electron system weakens. The existence of the low-energy branches, similar to the case of a single electron, supports the interpretation of the interaction between the 2D electron layer and the QD in terms of quasihole excitations of the Pfaffian state.

The charge distribution profile shown in Fig.\ \ref{fig_1e1h_charge} differs from the single-electron case shown in Fig.\ \ref{fig_1e_charge}. While the change in the system's angular momentum is not directly correlated with the shift of the distribution maximum in either case, the evolution of the charge distribution with $m$ varies between the one-electron and exciton QD systems.Specifically, for the one-electron system [Fig.\ \ref{fig_1e_charge}], variations occur primarily near the edge of the QD ($r \approx 2 l_0$) as the angular momentum increases, while the distribution near the origin and far from the QD remains unaffected. In contrast, for a QD exciton [Fig.\ \ref{fig_1e1h_charge}] increasing $m$ causes the charge distribution to change mainly near the origin (inside the QD), while it remains almost constant near the QD edge. This difference in behavior stems from the nature of the interaction with the 2D electron system: the interaction is of Coulomb type for a single electron, whereas it is of dipole type for an exciton system.  Interestingly, in Ref. \cite{Anyons_trions_PRX_2026}, the authors have proposed that the anyon-trion (i.e., an exciton bound to a quasihole) binding energy could be related to the fractional charge of the Laughlin quasihole, and therefore an important quantity for future experimental pursuit.

For the Laughlin state $(\nu=\frac13)$, excitons and charged excitons form in the QH system, as the electron density is rather low. We interpreted the anomaly of the cherged exciton energy to be due to creation of a fractionally-charged quasihole by the interaction with a localized charged exciton\cite{QD_FQHE_PRL_2003}. In the present case of Pfaffian quasiholes, a quantum well in close proximity to a correlated electron system conatining the $\frac52$ state, could perhaps provide the experimental setup. The charged particles, or excitons in the mionolayer could then show anomalies in energy and/or intensity at the fractional filling factor of $\frac52$. A somewhat similar method has been applied to 2D Transition Metal Dichalcogenide (TMDC) systems\cite{Correlated_insulating_states_Nature_2020,
Schuller}

Our theoretical approach where a QD containing a few electrons (or holes) Coulomb coupled to an incompressible Laughlin state, has been  established to provide unique insights on the nature of fractionally-charged excitations, in particular, the energetics of the quasiholes in the Laughlin state\cite{QD_FQHE_PRL_2003,QD_FQHE_exp_PhysicaE_2004,
charged_excitons_2004}. In the present case of the Pfaffian-Laughlin state, we observe clear gapped excitations with distinct branches at low momenta, that could be attributed to Pfaffian quasiholes. The splitting of the branches is caused by quasihole-quasihole interactions. The charge density distribution is consistent with the nature of the energy spectra. For an exciton in the QD, we also observe the low-energy branches of quasihole excitations, indicating the interactions between the Pfaffian state and the QD. Photoluminescene spectroscopy has been very successful in confirming our studies of the Laughlin state\cite{QD_FQHE_PRL_2003,QD_FQHE_exp_PhysicaE_2004,
charged_excitons_2004}. We expect that similar experimental studies of the Pfaffian state will illuminate the nature of the Pfaffian quasiholes. We have also evaluated the Pfaffian quasihole energy and found it to be in the range of Laughlin quasiholes.

The funding was provided by by Grant No. DE-SC0007043
from the Materials Sciences and Engineering Division of
the Office of the Basic Energy Sciences, Office of Science,
US Department of Energy. T.C. would like to thank Christian Sch\"uller (Regensburg, Germany) for valuable discussions.

%

\end{document}